\documentclass[12pt,preprint]{aastex}
\usepackage{float,epsfig}

\newbox\grsign \setbox\grsign=\hbox{$>$}
\newdimen\grdimen \grdimen=\ht\grsign
\newbox\laxbox \newbox\gaxbox
\setbox\gaxbox=\hbox{\raise.5ex\hbox{$>$}\llap
     {\lower.5ex\hbox{$\sim$}}}\ht1=\grdimen\dp1=0pt
\setbox\laxbox=\hbox{\raise.5ex\hbox{$<$}\llap
     {\lower.5ex\hbox{$\sim$}}}\ht2=\grdimen\dp2=0pt

\newcommand{\cha}{{\sl Chandra}}

\newcommand{\fer}{{\sl Fermi}}

\newcommand{\sax}{{\sl Beppo}-SAX}
\newcommand{\xte}{{\sl Rossi}-XTE}

\newcommand{\nsh}{$10^{21}\; {\rm cm^{-2}}$}

\begin{document}

\title{
{\sl Chandra} Phase-Resolved X-ray Spectroscopy of the Crab Pulsar II}

\author{
Martin C. Weisskopf \altaffilmark{1}, Allyn F. Tennant
\altaffilmark{1}, Dmitry G. Yakovlev \altaffilmark{2}, Alice Harding
\altaffilmark{3}, Vyacheslav E. Zavlin \altaffilmark{4}, Stephen L.
O'Dell \altaffilmark{1}, Ronald F. Elsner \altaffilmark{1}, and
Werner Becker \altaffilmark{5} }
\altaffiltext{1} {Space Sciences Department, NASA Marshall Space
Flight Center, VP62, Huntsville, AL 35812}
\altaffiltext{2} {Ioffe Physical Technical Institute,
Politechnischeskaya 26, 194021 St. Petersburg, Russia \& St.
Petersburg Polytechnical University, Politechnischeskaya 29, 195251
St. Petersburg, Russia}
\altaffiltext{3} {NASA Goddard Space Flight Center, 8800, College
Park, MD 20771}
\altaffiltext{4} {Universities Space Research Association, NASA
Marshall Space Flight Center, VP62, Huntsville, AL 35812}
\altaffiltext{5} {Max Planck Institut f\"ur Extraterrestrische
Physik, 85740 Garching bei M\"unchen, Germany}

\begin{abstract}
We present a new study of the X-ray spectral
properties of the Crab Pulsar. The superb angular resolution of the
\cha\ X-ray Observatory enables distinguishing the pulsar from the
surrounding nebulosity. Analysis of the spectrum as a function of
pulse phase allows the least-biased measure of interstellar X-ray
extinction due primarily to photoelectric absorption and secondarily
to scattering by dust grains in the direction of the Crab Nebula. We
modify previous findings that the line-of-sight to the Crab is
under-abundant in oxygen and provide measurements with improved
accuracy and less bias. Using the abundances and cross sections from
Wilms, Allen \& McCray (2000) we find [O/H] = $(5.28 \pm
0.28)\times10^{-4}$ ($4.9 \times10^{-4}$ is solar abundance).
We also measure for the first time the impact of
scattering of flux out of the image by interstellar grains. We find
$\tau_{\rm scat} = 0.147 \pm 0.043$. Analysis of the spectrum as a
function of pulse phase also measures the X-ray spectral index even
at pulse minimum~--- albeit with increasing statistical uncertainty.
The spectral variations are, by and large, consistent
with a sinusoidal variation. The only significant variation from the
sinusoid occurs over the same phase range as some rather abrupt
behavior in the optical polarization magnitude and position angle. We
compare these spectral variations to those observed in Gamma-rays and
conclude that our measurements are both a challenge and a guide to
future modeling and will thus eventually help us understand pair
cascade processes in pulsar magnetospheres. The data were also used
to set new, and less biased, upper limits to the surface temperature
of the neutron star for different models of the neutron star
atmosphere.
We discuss how such data are best
connected to theoretical models of neutron star cooling and neutron
star interiors. The data restrict the neutrino emission rate in the
pulsar core and the amount of light elements in the heat-blanketing
envelope. The observations allow the pulsar, irrespective of the
composition of its envelope, to have a neutrino emission rate higher
than 1/6 of the standard rate of a non-superfluid star
cooling via the modified Urca process. The observations also allow the
rate to be lower but now with a limited amount of accreted matter in
the envelope.
\end{abstract}


\section{Introduction} \label{s:intro}

In Weisskopf et al.\ (2004) --- hereafter Paper I--- we presented the
first \cha-LETGS (Low Energy Transmission Grating Spectrometer)
phase-averaged and phase-resolved X-ray spectroscopy of the Crab
Pulsar. In that paper we set an upper limit to the thermal
emission from the neutron star's surface, essentially unblemished by
any contaminating signal from the pulsar's wind nebula. Here we
present the results of new \cha\ observations that made use of a High
Resolution Camera (HRC) shutter. The HRC serves as the readout for
the LETGS. Using the shutter, together with a factor of two for
increased observing time, allows for more high-time resolution data
over our previous observation by an order of magnitude and thus more
meaningful phase-resolved spectroscopy.

After briefly describing the observation and data reduction (\S
\ref{s:obs}), we discuss the analysis of the measured spectra (\S
\ref{s:anares}). We update the work of Paper I regarding the
interstellar abundances in the line of sight to the pulsar and other
relevant parameters (\S \ref{s:pas}). We do not repeat the discussion
in Paper I concerning the impacts on the spectroscopy from using
different scattering coefficients and cross-sections, nor do we
repeat the comparison of our results with certain other previous
measurements. We present (\S \ref{s:svar}) new, more precise,
measurements of the variation of the non-thermal spectral parameters
with pulse phase and the implications. We discuss constraints on the
surface temperature of the underlying neutron star (\S \ref{s:temp})
assuming two different models for the thermal emission and  fitting
the data as a function of pulse phase allowing the power law
component to vary. This approach yields a less-biased approach to
measuring, or setting upper limits to, the thermal spectral
parameters. We discuss ramifications of the temperature measurements
(\S \ref{s:imp}) and summarize our findings (\S \ref{s:sum}).

\section{Observations and Data Reduction} \label{s:obs}

In this paper we make use of data (ObsID 9765) we obtained in 2008,
January. As with our previous observation (Paper I) these data were
taken using \cha 's Low-Energy Transmission Grating (LETG) and
High-Resolution Camera spectroscopy detector (HRC-S)~--- the LETGS.
For this observation, however, we also inserted one of the HRC's
shutters to occult most of the positive order, a significant fraction
of the zero order, and some of the negative order flux. This was done
deliberately to prevent the total counting rate from exceeding the
telemetry limit\footnote{{\tt http://asc.harvard.edu/proposer/POG/}}. Not exceeding the telemetry limit, in turn, allows
us to make full use of the HRC time resolution without having to
employ the severe filtering described in Paper I which dramatically
reduced the amount of detected events with time resolution good
enough to resolve the light curve of the pulsar. In addition, and
also to keep the telemetry rate low, the two outer HRC-S detectors
were turned off.

We processed all data using \cha\ X-ray Center (CXC) tools. Level 2
event files were created using the CIAO script {\tt hrc\_process\_events} with pixel randomization off and CALDB 4.4.1. The program
{\tt axbary} was used to covert the time of arrival of events to the
solar system's barycenter. Using LEXTRCT (developed by one of us,
AFT) we extracted the pulsar's dispersed spectrum from the images.
The extraction used a $\pm12.348$-pixel-wide (cross-dispersive)
region centered on the pulsar, in 2-pixel (dispersive) increments.
This extraction width is the standard extraction region for which the
CIAO-derived response functions apply for the central HRC-S detector.
For the LETG's 0.9912-$\mu$m grating period and 8.638-m
Rowland-circle radius, the LETGS dispersion is 1.148 \AA /mm.
Consequently, the spectral resolution of the binned data (two
6.4294-$\mu$m HRC-S pixels) is 0.01476 \AA . Because of the
occultation of the HRC shutter, we only considered the minus-order
dispersed spectrum.

The occultation is a geometrical effect with the shutter occulting
progressively less and less of the negative order with the degree of
obscuration varying from about 75\% at 3.8 keV to only about 10\% at
0.3 keV. The response function is the standard CIAO response but
applying a correction factor simply determined from the ratio of the
minus-order ObsID 759 (Paper I) flux to that seen in the ObsID 9765
data to account for the effect of shuttering.

We restricted all data analysis to the (first-order) energy range 0.3
to 3.8 keV. 
The upper spectral limit avoids contamination from the zeroth-order nebular image (see Fig. 1 in Paper I); the lower limit is placed at a level where the first order flux effectively goes to zero.

Selecting the appropriate region of the data to be used as the
background estimator requires some care as one must deal with the
dispersed flux from the nebula which includes such bright features as
the inner ring (Weisskopf et al.\ 2000). To determine the background
we studied the flux at the deepest portion of pulse minimum (phase
range $0.65$ to $0.70$, Fig.~\ref{f:3}) in the spectral region
corresponding to 0.3 to 3.8 keV in first order and projected onto the
cross-dispersion axis (Fig.~\ref{f:back}). We chose pulse minimum for
this study to avoid having the pulsar dominate the projected data.
Based on the information shown in Fig.~\ref{f:back}, and to estimate
the background, we extract data from two 15-HRC-pixel-wide bands
starting 15 pixels to either side of the pulsar's dispersed image.
Note that just beyond (more negative)
$-15$ pixels in Fig.~\ref{f:back} there is a slight rise traceable to
the dispersed spectrum of features near the pulsar, showing that
extending the background region much wider would be a mistake. Of
course the asymmetric nature of the background in the
cross-dispersion direction due to features in the nebula makes the
selection of regions to either side of the dispersed spectrum a
necessity, so that the average provides an accurate background
estimate.

\clearpage

\vspace{10pt}
\begin{center}
\includegraphics[angle=-90,width=\columnwidth]{fig1_flux_vs_pixels_300eV.ps}
\vspace{10pt} \figcaption{The projected flux in the negative order
from 0.3 to 3.8 keV versus position in the cross-dispersion
direction. These data cover the pulse-phase range 0.65$-$0.70. The two
vertical lines indicate the regions selected for background
estimation. The sloping line (the fit to the background) indicates
that this choice of regions symmetrically, and accurately, estimates
the underlying background in the dispersed pulsar spectrum.
\label{f:back} }
\end{center}

\clearpage

\section{Analysis and Results} \label{s:anares}

As in Paper I, whenever we fit data to a particular spectral model we
allow for interstellar absorption using cross-sections from {\tt
vern}, Verner et al.\ (1996), and interstellar absorption {\tt
tbvarabs}, Wilms, Allen \& McCray (2000, references therein), allowing for
absorption by interstellar grains. In addition, owing to the small
effective aperture of our observation of the pulsar, we also allow
for the effects of diffractive scattering by grains on the
interstellar extinction using a formula based on the Rayleigh--Gans
approximation (van der Hulst 1957; Overbeck 1965; Hayakawa 1970),
valid when the phase shift through a grain diameter is small. Details
may be found in Paper I. We analyzed the data using the XSPEC
(v.11.3.2) spectral-fitting package (Arnaud 1996). To ensure
applicability of the $\chi^{2}$ statistic, we merged spectral bins as
needed to obtain at least 100 counts per fitting bin (before
background subtraction). Response functions were first generated
using the CIAO threads {\tt mkgarf} and {\tt mkgrmf} and then
corrected for the occultation of the shutter blade. The effective
areas were generated using standard CIAO tools and CALDB 4.4.3

\subsection{Phase-Averaged Spectrum} \label{s:pas}

We present here the results of our analysis of the phase-averaged
spectrum, although we will not use this spectrum to search for an
additional thermal component nor to measure phase-independent parameters. These tasks are more appropriately
accomplished using the phase-resolved spectrum
(\S~\ref{s:svar}), a point also emphasized by Jackson \& Halpern
(2005).

Fitting the phase-averaged data to a power law spectrum and allowing
for interstellar absorption, a variation in the abundance ratio of
oxygen to hydrogen, and small-angle scattering by intervening dust
give an excellent fit, $\chi^{2}$ of $1795$ for $1810$
degrees-of-freedom ($\nu$). 
The best-fit spectrum is shown in Fig.~\ref{f:spectrum}.

In Table~\ref{t:phavg} we list the best-fit parameters, but now fix the dust scattering factoring as was done in Paper I to make comparisons.
As a convenience for the reader, the first line in the table repeats the results from Paper I.
The second line shows how these results change using the more recent response function, background extraction regions, etc.
The reader will note that there are differences between line 1 and 2 of the Table that are beginning to approach statistical significance, primarily the change in the hydrogen abundance is lower (3.68 as compared to 4.20).
The differences are mostly a consequence of the evolution of the calibration of the LETGS effective area.
The differences between the results for the same data (lines 1 and 2 of  Table~\ref{t:phavg}) emphasize that we have ignored (as many X-ray astronomers do) the possibility that there are any errors in the response functions! 
The final comparison is between our current results (line 3 of Table~\ref{t:phavg}) and the updated results for ObsID 759. 
As before, there are differences, but none of them at the 3$\sigma$ level. 
The largest differences, in the normalization and in the powerlaw index may be a reflection of inaccuracies at the 5\%-level in our correction to the response function for the insertion of the blade.

\vspace{10pt}
\begin{center}
\includegraphics[angle=-90,width=\columnwidth]{fig_phase_avg_spectrum.ps}
\vspace{10pt} \figcaption{Fit of the phase-averaged data to a power
law model allowing 5 variables: the power law index, the
normalization, $N_{\rm H}$, the oxygen to hydrogen ratio $[O/H]$, and the dust-scattering constant. The upper curves compare the data to
the best-fit model folded through the response function. The lower
curves show the contributions to $\chi^2$. The plotted data were
combined into larger bins for visual clarity.
\label{f:spectrum} }
\end{center}

\begin{table}
\begin{center}
{\caption {Power law fit$^{a}$ to the \cha-LETGS phase-averaged spectrum of the
Crab Pulsar in the band from 0.3 to 3.8 keV holding the dust-scattering factor constant at 0.15. \label{t:phavg}}}
\begin{tabular}{lllllll}
\hline
ObsID         & $\chi^{2}$/$\nu$& $\Gamma_{\rm P}$   & $N_{\rm H}$  & $[O/H]$    & $norm$         \\
              &                 &                    & \nsh         &$10^{-4}$   &                \\ \hline
Paper I$^{b}$ & $1539/1552^{c}$ & $1.587\pm0.019^{d}$    &$4.20\pm0.14$&$3.33\pm0.44$&$0.506\pm0.008$ \\ \hline
759$^{e}$     & $1447/1476$     & $1.622\pm0.023$    &$3.68\pm0.13$&$4.52\pm0.42$&$0.479\pm0.009$ \\ \hline
9765          & $1832/1811$     & $1.538\pm0.023$    &$3.59\pm0.11$&$3.84\pm0.32$&$0.450\pm0.008$ \\ \hline
\end{tabular}
\tablecomments{\\
$^{a}$ Abundance models in XSPEC: {\tt wilm}, Wilms, Allen \& McCray (2000); Cross-section models in XSPEC: {\tt vern}, Verner et al.\ (1996); {\tt tbvarabs}, Wilms, Allen \& McCray (2000, references therein) allowing
for absorption by interstellar grains.\\
$^{b}$ ObsID 759 (Paper I).\\ 
$^{c}$ Larger extraction width, hence more counts, hence larger number of degrees of freedom compared to line 2. \\
$^{d}$ Uncertainties are XSPEC's estimate of the $1-\sigma$ error treating
each variable as the one interesting parameter ($\chi^{2}$ at minimum $+ 1.0$).\\
$^{e}$ Paper I, but using the up-to-date response function, and extraction widths. \\
}
\end{center}
\end{table}

\clearpage

\subsection{Spectral Variation with Pulse Phase} \label{s:svar}

We were previously limited in our ability to study the spectrum as a
function of pulse phase (ObsID 759 and Paper I) because a HRC-S timing
error assigns to each event the time of the previous event, thus
complicating the analysis for this bright source when telemetry is
saturated and events are dropped. In Tennant et al. (2001) we discussed this
problem and a method for maintaining some timing accuracy under these
conditions~--- albeit at significantly reduced efficiency. 
The method filters the data, accepting
only telemetered events separated by no more than 2 ms, guaranteeing
a timing accuracy never worse than 2 ms and typically much better.
Thus, although there were approximately 50 ksec of observing time for
ObsID 759, only a small fraction (1/15) of the data were useful for
studying spectral variations with pulse phase. For the
current (ObsID 9765) observation, which lasted for about 100 ksec,
this problem does not exist since the observation was designed so
that the telemetry never saturated.

Jodrell Bank (Lyne, Pritchard, \& Smith 1993) routinely observe the
Crab Pulsar (Wong, Backer \& Lyne 2001) providing a period ephemeris
\footnote{{\tt http://www.jb.man.ac.uk/$\sim$pulsar/crab.html}}.
Roberts \& Kramer (2000, 2008, private communications) kindly prepared
ephemerides matched to our observation times.

In performing the phase-resolved spectral analysis, we allow the
interstellar absorption and dust scattering parameters to vary, but
assume that they are identical for each pulse phase bin. However, the spectral index and normalization is allowed to vary. This removes
any bias produced by assuming a power law index that is independent
of phase. The fit to the data was excellent, $\chi^2$ was 3156 on
3207 degrees of freedom. The best-fit values for the
non-phase-varying parameters and their approximate
(one-interesting-parameter) uncertainties are given in
Table~\ref{t:varyphase}. Note $\tau_{\rm scat}$, which we previously
(Paper I) postulated must be present and accounted for, has now been
detected at almost $4\sigma$. Table~\ref{t:phase} and
Figure~\ref{f:3} summarize the results for the pulsar's
phase-resolved power law photon index, $\Gamma_{\rm P}$.

\begin{table}
\begin{center}
{\caption {Power law fit$^{a}$ to the \cha-LETGS phase-resolved spectrum of the Crab Pulsar. \label{t:varyphase}}}
\begin{tabular}{lcccc}
\hline
ObsID  & $\chi^{2}$/$\nu$ & $N_{\rm H}$    & $[O/H]$        &$\tau_{\rm scat}$ \\
      &                  & \nsh           & $10^{-4}$      &$ @1keV$          \\ \hline
9765  & $3510/3546$      & $3.22\pm0.12^b$& $5.28\pm0.28^c$&$0.147\pm0.043^c$ \\ \hline
\end{tabular}
\tablecomments{\\
$^{a}$ Abundance models in XSPEC: {\tt wilm}, Wilms, Allen \& McCray (2000); Cross-section models in XSPEC: {\tt vern}, Verner et al.\ (1996);
{\tt tbvarabs}, Wilms, Allen \& McCray\ (2000, references therein) allowing for
absorption by interstellar grains.\\
$^{b}$ Uncertainties are XSPEC's estimate of the $1\sigma$ error treating
each variable as the one interesting parameter ($\chi^{2}$ at minimum $+ 1.0$).\\
\\
}
\end{center}
\end{table}

\clearpage

\begin{table}
\begin{center}
\caption {Power law Index versus Pulse Phase.
\label{t:phase}}
\begin{tabular}{cc}
\hline
Phase Range $\times 1000$ & $\Gamma_{\rm P}$  \\
\hline
001$-$017  & $ 1.627 \pm 0.041^{a} $ \\
017$-$031  & $ 1.582 \pm 0.063 $ \\
031$-$051  & $ 1.462 \pm 0.073 $ \\
051$-$075  & $ 1.461 \pm 0.088 $ \\
075$-$120  & $ 1.449 \pm 0.097 $ \\
120$-$330  & $ 1.341 \pm 0.040 $ \\
330$-$370  & $ 1.471 \pm 0.037 $ \\
370$-$390  & $ 1.499 \pm 0.039 $ \\
390$-$400  & $ 1.535 \pm 0.047 $ \\
400$-$410  & $ 1.621 \pm 0.051 $ \\
410$-$430  & $ 1.604 \pm 0.050 $ \\
430$-$470  & $ 1.655 \pm 0.059 $ \\
470$-$650  & $ 1.698 \pm 0.010 $ \\
650$-$830  & $ 1.886 \pm 0.046 $ \\
830$-$950  & $ 1.502 \pm 0.058 $ \\
950$-$960  & $ 1.733 \pm 0.060 $ \\
960$-$970  & $ 1.611 \pm 0.049 $ \\
970$-$980  & $ 1.649 \pm 0.042 $ \\
980$-$984  & $ 1.651 \pm 0.052 $ \\
984$-$992  & $ 1.627 \pm 0.039 $ \\
992$-$001  & $ 1.594 \pm 0.039 $ \\
\hline
\end{tabular}
\tablecomments{\\
$^{a}$ Uncertainties are XSPEC's estimate of the $1\sigma$ error treating
each variable as the one interesting parameter ($\chi^{2}$ at minimum $+ 1.0$).\\
\\
}
\end{center}
\end{table}

\clearpage

Based upon a $\chi^2$ analysis of the distribution of best-fit photon
indices (Table~\ref{t:phase},  Figure~\ref{f:3}), we reject, with
high confidence, the hypothesis that the spectral index is constant
with phase. The error-weighted average of the spectral indices is
1.563 and the value of $\chi^2$ was 71 on 20 degrees of freedom. The
variation of spectral index between pulse phases -$0.1$ and $0.5$ is
qualitatively similar in \cha\ (Paper I, this paper), \sax\ (Massaro
et al.\ 2000), and \xte\ (Pravdo, Angelini, \& Harding 1997)
measurements, with the index increasing (becoming softer) through the
primary-pulse maximum and decreasing (becoming harder) in the bridge
between the primary and secondary pulses. It is difficult to be more
quantitative in this comparison as the non-\cha\ data were analyzed
using different cross-sections, different abundances, and covered
different spectral ranges. Moreover, \cha\ provides the angular
resolution needed to isolate the pulsar from the nebula, something
that is essential to measure the spectral index for the pulse-phase
range 0.5--0.9. Our analysis shows that the spectral index at pulse
minimum is consistent with an apparent continuation of the increase
(softening) of the spectral index until just before the onset of the
primary pulse. The spectral-index uncertainty near pulse minimum is,
of course, large because there are fewer counts. It is also
interesting that the slope of the spectral variations appears to be
the highest (softest) during the peak of the two pulses.

In Fig.~\ref{f:3} we also show the results of fitting a constant plus
a sine wave to the variation of the spectral index but excluding the
three data points with the largest uncertainty, i.e. those in the
phase range from $0.47$ to $0.95$ --- pulse minimum and the points
just before and just after pulse minimum. The fit to the sine wave is
excellent and is probably the type of behavior one might expect as
the orientation of the pulsar's magnetic field varies with phase to
the distant observer. The phase of the variation is (probably) an
indication of the geometry. The point just before the rise to the
pulse maximum, however, clearly does not fit this simple picture.

\clearpage

\vspace{10pt}
\begin{center}
\includegraphics[angle=-90,width=\columnwidth]{fig_plindex_lc_P_Pa_vs_phase.ps}
\vspace{10pt} \figcaption{The upper curve shows the measured
variation of the powerlaw index. The solid curve is the result of
fitting a sine wave plus a constant to these data not including the
three points in and around pulse minimum which exhibit the largest
error in the powerlaw index. The X-ray light curve (background not subtracted) is shown in the second panel from the top. The bottom two panels show the variation
of the optical degree of polarization and position angle. The optical
data are from Slowikowska et al.\ (2008). Slight differences, if any,
between the optical pulse phase and the X-ray pulse phase have been
ignored. One pulse cycle is repeated twice for clarity.
 \label{f:3}}
\end{center}

\clearpage

\subsubsection{Discussion\label{s:disc}}
An interesting correlation is shown also in Fig.~\ref{f:3} in the
bottom two panels which plot optical polarization data kindly
provided by G.~Kanbach and A.~Slowikowska (Slowikowska et al.\ 2008).
Here we emphasize the afore-mentioned change of behavior in the X-ray
power law index just before the rise of the light curve to primary
pulse maximum (phases 0.83 to 0.95) and the abrupt change of optical
polarization and position angle in this same phase interval. These
would appear to be correlated phenomena.

The variation of spectral index with phase shown in Fig.~\ref{f:3}
and Table~\ref{t:phase} is also strikingly similar to the spectral
index variations measured by the \fer\ Gamma-Ray Space Telescope
above 100 MeV (Abdo et al.\ 2010a), with the hardest (smallest) index
occurring midway between the two peaks and rising symmetrically
through both peaks to reach maxima in the off-peak region. There is
also even a hint in the Fermi data of the small maximum preceding the
first peak. Indeed, the photon index variation is similar in other
bright gamma-ray pulsars, including Geminga (Abdo et al.\ 2010b) and
Vela (Abdo et al.\ 2010c), where the maximum preceding the first peak
is even more pronounced. Yet, the Crab broadband spectrum is very
different from that of Vela or Geminga, being one of very few pulsars
(that include B1509$-$58) having equal or greater power in the X-ray
band as in the hard gamma-ray band. The multiwavelength spectrum of
the Crab pulsar (Kuiper et al. 2001) seems to comprise two distinct
components: one extending from UV to soft gamma-rays and one
extending from soft to hard gamma-rays (although the position of the
division varies somewhat with pulse phase). In the phase-resolved
spectra, the spectral indices of the two components tend to mirror
each other, with the hardest spectra in soft X-rays and hard
gamma-rays occurring in the bridge region and the softer spectra
occurring in the peaks. The similarity of the \cha\ and \fer\
spectral index behavior is consistent with this trend.

The fact that the soft X-ray and hard gamma-ray spectra are part of
two seemingly different radiation components, and most likely have
different emission mechanisms, raises the question of why their
spectral index variation with phase should be so similar. They must
share a common property, such as the same radiating particles or the
same locations in the magnetosphere.

It is now widely agreed that the high energy emission from pulsars
originates in their outer magnetospheres, since the measurement of
the cutoffs in their gamma-ray spectra rules out attenuation due to
magnetic pair production and therefore emission near the polar caps
(Abdo et al.\ 2009). Several different outer-magnetosphere models,
that advocate different emission mechanisms in the X-ray range, make
predictions for phase-resolved spectral variations. In outer gap
models (Cheng, Ho \& Ruderman 1986, Romani 1996), particles are
accelerated in vacuum gaps that form along the last open magnetic
field lines, from above the null-charge surface to near the light
cylinder. In slot gap  models (Muslimov \& Harding 2004), particles
are also accelerated along the last open field lines, but in a
charge-depleted layer from the neutron star surface to near the light
cylinder. In both models, the high-energy peaks in the light curve
are caustics, caused by cancellation of phase differences along the
trailing field lines (Morini 1983) or overlapping field lines near
the light cylinder.

Harding et al.\ (2008) presented a model for 3D acceleration and
high-altitude radiation from the slot gap, with application to the
Crab pulsar. In this model, emission in the optical to soft gamma-ray
band is synchrotron radiation from pairs outside the slot gap
undergoing cyclotron resonant absorption of radio photons. Hard
gamma-rays come from primary electrons accelerating in the slot gap
and radiating curvature and synchrotron emission. The common element
for the X-ray and gamma-ray emission would then be the angles to the
radio photons. Since Harding et al.\ (2008) assumed that the pair
spectrum was constant throughout the open field volume, there was no
spectral index variation with phase. However, polar cap pair cascade
simulations (the origin of the X-ray emitting pairs in this model)
show that there are large variations in the pair spectrum across the
polar cap (Arendt \& Eilek 2002). Thus the detailed measurements of
X-ray spectral index variation presented in this paper is mapping
(and constraining) the variation in the pair spectrum across the open
field lines.

In recent studies of phase-resolved spectra of the Crab pulsar in the
outer gap model (Tang et al.\ 2008, Hirotani 2008), the optical to
hard X-ray spectrum comes from synchrotron radiation of secondary
pairs produced in situ in outer gap cascades while the gamma-rays
come from inverse Compton radiation of pairs and curvature radiation
of primary particles. This model does not match the observed X-ray
spectral variations, although it does somewhat produce the observed
gamma-ray spectral variations. Thus, this model lacks the essential
physics that accounts for the X-ray spectral index variation and its
similarity to that in gamma-rays.

\clearpage

\subsection{Temperature of the Neutron Star and Superfluidity}\label{s:temp}

Here we investigate the hypothesis that there is a detectable
underlying thermal component in addition to the non-thermal flux that
we see from the pulsar. We fit the data as a function of pulse phase
to spectral models that allow both components. We first consider a
power law together with a thermal black-body and then a model with a
spectrum of radiation emergent from the hydrogen neutron star
atmosphere. In XSPEC these models are the {\tt powerlaw}, {\tt
bbodyrad}, and {\tt nsa} (Pavlov et al.\ 1995).

We examined two approaches for the analysis. In the first, we use the data from all 21 pulse phase bins, in the second, we use the data from the 4 phase bins that are at pulse minimum. In the latter case we use the data from the other 17 phase bins to establish the values of the phase independent parameters $N_{\rm H}$, $[O/H]$, and  $\tau_{\rm scat}$ for fitting the data at pulse minimum. We found that the sensitivity to a thermal component was virtually the same in both approaches, however, the establishment of upper limits was computationally much faster using the data at pulse minimum. 

\subsubsection{Blackbody Model} \label{s:bb}

Adding a phase-independent black-body model to the spectral fitting
yields the results listed in Table~\ref{t:bbody}.
The large uncertainties in both the normalization,
$\theta_{\infty}^2$, and the redshifted effective surface
temperature, $kT_{\infty}$, clearly point to the absence of a
blackbody component within statistics. $\theta_{\infty}$ is the
angular size determined by a distant observer, in XSPEC units~---
$\theta_{\infty} = (R_{\infty}/D_{10})$, with $R_{\infty}$ the
apparent stellar radius in km units and $D_{10}$ the source distance
in 10-kpc units. Fig.~\ref{f:kT_bbnorm} shows the 2- and 3-$\sigma$
upper limits to $kT_{\infty}$ for a range of values for
$\theta_{\infty}^2$ that are relevant to neutron star models with
realistic equations of state.

In Paper I we found a somewhat lower 3-$\sigma$ upper limit and a higher 2-$\sigma$ upper limit to $kT_{\infty}$ than those shown in Figure~\ref{f:kT_bbnorm}.
However, these results were erroneous and should have been higher. Unfortunately this error was only discovered while completing this paper.
As an example of the changes due to the error, the 2- and 3-$\sigma$ upper limits at $\theta_{\infty}^{2} = 6100$ should have been 0.195 and 0.209 keV respectively. 
Using the newer response function and signal and background extraction regions would have lowered these to 0.184 and 0.202 keV respectively.
Thus, the upper limits reported here, as expected, represent a significant improvement over Paper I.
(We note again that the difference in upper limits using the old and new response functions ignores the possibility that there are uncertainties associated with these responses.) 
Of course here we analyze the data as a function of pulse phase.
As discussed above, this is a better approach as it removes any
possible bias produced by averaging a number of power laws, which, in
turn won't be a powerlaw and thus inadvertently create a spurious
thermal component. 

\begin{table*}
\begin{center}
\caption {Best-fit values for the phase-independent parameters after analyzing the phase-resolved data using a {\tt bbodyrad} plus {\tt powerlaw} model. The latter is allowed to vary as a function of phase.
Uncertainties are XSPEC estimates for the 1$\sigma$ statistical errors based on one interesting parameter ($\chi^{2}$ at minimum +1.0).\label{t:bbody}}
\begin{tabular}{cc}
\hline
Parameter   &                      \\
\hline
$\chi^{2}/\nu$ &   $3510/3544$        \\
$N_{\rm H} (10^{21} cm^{-2})$ & $3.22\pm 0.13$  \\
$[O/H]$  & $(4.28 \pm 0.30)\times10^{-4}$ \\
$\tau_{\rm scat}\ at\  1 keV $ &  $0.147 \pm 0.045$\\
$kT_{\infty}(keV)$ & $ 0.1 \pm 7.2 $ \\
$\theta_{\infty}^{2}$ & $44 \pm 31000$ \\
\hline
\end{tabular}
\end{center}
\end{table*}

\vspace{10pt}
\begin{center}
\includegraphics[angle=-90,width=\columnwidth]{fig_kT_vs_thsq_bb.ps}
\vspace{10pt}
\figcaption {The 2- and 3$-\sigma$ upper limits to $kT_{\infty}$ derived (lower and upper curves respectively) using the {\tt powerlaw}+{\tt bbodyrad} spectral model. The limits are based on a single interesting parameter ($\Delta\chi^{2} = \chi^{2}-\chi^{2}_{\rm min} = 4.0$ (lower curve) and $9.0$ (upper curve) and plotted as a function of $\theta_{\infty}^{2}$.
(See text for details.)
 \label{f:kT_bbnorm}}
\end{center}

\subsubsection{NSA Model} \label{s:nsa}

The {\tt NSA} model requires a number of inputs. The normalization
was set assuming a distance to the Crab of 2 kpc. The surface
magnetic-field parameter was set at $1.0\times10^{13}$ Gauss, although results are not terribly sensitive to this choice.
The model also requires the gravitational mass
$M$ and the circumferential radius $R$ of the neutron star.
We examined a wide range of $M$ from 1.0 to 2.5 $M_\odot$ and $R$
from 8 to 15 km in creating Fig.~\ref{f:kT_nsa} which shows the 2- and
3-$\sigma$ upper limits to $kT_\infty$; hence the reason for the
multiple values for a given $\theta_{\infty}$.

\vspace{10pt}
\begin{center}
\includegraphics[angle=-90,width=\columnwidth]{fig_kT_vs_thsq_nsa.ps}
\vspace{10pt}
\figcaption {The 2- and 3$-\sigma$ upper limits to $kT_{\infty}$ (lower and upper curves respectively) fitting a {\tt powerlaw}+{\tt NSA} model. The limits are based on a single interesting parameter ($\Delta\chi^{2} = \chi^{2}-\chi^{2}_{\rm min} = 4.0$ (lower curve) and $9.0$ (upper curve) as a function of $\theta_{\infty}^{2}$. Multiple values of $kT_{\infty}$ arise as different combinations of $M$ and $R$ lead to the same (or similar) values of
$\theta_{\infty}$.
 \label{f:kT_nsa}}
\end{center}
\clearpage

The reader will note that {\tt powerlaw}+{\tt NSA} fits always yield
a higher upper limit for given values of $\theta_{\infty}$ than {\tt powerlaw}+{\tt bbodyrad}.
Simulations with fake data have shown this to be correct. We believe
that this happens because the NSA spectrum has a hard tail which makes it difficult to distinguish from the power law.

\subsubsection{Implications} \label{s:imp}

Here we apply cooling theory for neutron stars and formulate
constraints on the internal structure of the Crab pulsar that can be
inferred from the observational upper limits of its effective surface
temperature $T$
(\S \ref{s:bb} \& \S \ref{s:nsa}).

Current cooling theories (e.g.\  Page, Geppert \& Weber 2006, Page et
al.\ 2009, Yakovlev \& Pethick 2004 , Yakovlev et al.\ 2008) state
that any isolated neutron star of the Crab pulsar age ($t \sim 10^3$
yr) should be at the neutrino cooling stage and have an isothermal
interior. The preceding cooling stage of internal thermal relaxation,
when the neutron star core is noticeably colder than the crust
(because of stronger neutrino cooling of the core and slower thermal
conduction in the crust), lasts no longer than $\sim 200$ yrs. That
stage should be over. The interior of the pulsar then should be
highly isothermal having the same internal temperature
$\widetilde{T}_i(t)$,
where
$\widetilde{T}_i$
is the internal temperature redshifted for a distant observer
(Thorne, 1977). The local (actual) temperature
$T_i$
in the isothermal interior is $\sim 10-30$\% higher than
$\widetilde{T}_i$,
with the stellar core being somewhat hotter than the crust. A
noticeable temperature gradient in a thermally relaxed star survives
only near the surface, in the outer heat-blanketing envelope
(Gudmundsson, Pethick \& Epstein 1983) with thickness not higher than
a few tens of meters. In the envelope, the temperature drops from the
temperature inside the star to the effective surface temperature
$T$.
The temperature drop depends on the matter composition and on the
magnetic field in the envelope (which affects the thermal
conductivity -- see Potekhin, Chabrier, \& Yakovlev 1997, Potekhin et
al.\ 2003).

The cooling of the Crab pulsar (as of all isolated neutron stars of
ages $t \lesssim 10^5$ yrs) is driven by neutrino emission from its
interior, mainly from the superdense core. The decrease of the
internal temperature $\widetilde{T}_i(t)$ with time is determined by the physics of the core, being insensitive
at this cooling stage to the physics of the envelope. Therefore,
the internal cooling of a star with a given internal structure is the
same for any heat-blanketing envelope (looks the same from inside)
but the surface temperature is affected by the particular properties
of the envelope (looks different from outside).

There are numerous versions of current cooling theories as cited
above and they are still poorly constrained by observations. The
theories comprise different compositions and equations of state
(EOSs) of neutron star cores, different neutrino emission properties
and models for superfluidity of baryons (which affect heat capacity
and neutrino emission). In spite of the large number of scenarios,
the cooling of isolated thermally relaxed neutron stars with an
isothermal interior is mostly regulated by the three factors
\footnote{This statement is true as long as a noticeable temperature
decrease $T_s(t)$ is not observed for a given neutron star, which is
so for all currently observed isolated neutron stars except for the
star in the Cas~A supernova remnant (Heinke \& Ho 2010). In this case
one can glean more information as to the neutron star structure (Page
et al.\ 2011, Shternin et al.\ 2011).}
(e.g.  Yakovlev et al.\ 2011): (i) the neutrino cooling rate; (ii)
the properties of the heat-blanketing envelope; and (iii) the stellar
compactness.

The neutrino cooling rate $\ell$ [K s$^{-1}$] is defined as
\begin{equation}
    \ell=L_\nu^\infty(\widetilde{T}_i)/C(\widetilde{T}_i).
\label{ell}
\end{equation}
Here, $L^\infty_\nu(\widetilde{T}_i)$ is the neutrino luminosity of the star (redshifted for a distant observer), and $C(\widetilde{T_i})$ is the heat capacity (see, e.g.
Eqs.\ (3) and (5) in Yakovlev et al.\ 2011). It is instructive
(Yakovlev et al.\ 2011) to introduce the normalized cooling rate
\begin{equation}
     f_\ell=\ell(\widetilde{T}_i)/\ell_{\rm SC}(\widetilde{T}_i),
\label{fl}
\end{equation}
where $\ell_{\rm SC}(\widetilde{T}_i)$ is the neutrino cooling rate of the {\it standard neutrino candle}, a
neutron star with the same $M$, $R$, $\widetilde{T}_i$, but with a non-superfluid nucleon core that is cooling via the ordinary
modified Urca process of neutrino emission. For isolated neutron
stars without any additional internal heat sources, physically
allowable values of $f_\ell$ may vary from $\sim 10^{-2}$ to $\sim
10^6$ (e.g. Yakovlev et al.\ 2011). If $f_\ell \sim 1$ this implies
standard neutrino cooling, $f_\ell \sim 10^{-2}$ very slow cooling
(e.g.\  when the modified Urca process is suppressed by
superfluidity) and $f_\ell \sim 10^2-10^6$ fast cooling
(accelerated by direct Urca processes, pion or kaon condensates, or
by neutrino emission due to Cooper pairing of neutrons).

We have employed the models of heat-blanketing envelopes of Potekhin,
Chabrier, \& Yakovlev (1997) and Potekhin et al.\ (2003) which may
contain some mass $\Delta M$ of (light-element) accreted matter and
have a dipole magnetic field $B$ (with $B=3.8 \times 10^{12}$~G at
the magnetic equator for the Crab pulsar). The effective temperature
of a magnetized star varies over the surface with the magnetic poles
being hotter than the equator. Cooling theory (e.g.
Potekhin, Chabrier \& Yakovlev 1997) suggests one use the effective
temperature averaged over the surface (it defines the thermal
luminosity of the star). The effect of the given
magnetic field on the cooling is weak although included in our
calculations. The effect of the envelope on the cooling is regulated
then by $\Delta M$ which varies from $\Delta M=0$ for
a standard iron envelope to $\Delta M_\mathrm{max} \sim
10^{-7}\,M_\odot$ for a fully accreted envelope. Larger $\Delta M$
are not realistic because light elements transform into heavier ones
at the bottom of the envelope through electron capture and nuclear
reactions.

The compactness of the star can be characterized by the parameter
\begin{equation}
x=\frac{2GM}{Rc^2}\approx 2.95 \; \frac{M}{M_\odot} \; \frac{1~{\rm
km}}{R}
\label{eq:x}
\end{equation}
which is the ratio of the Schwarzschild radius to $R$. According to
Yakovlev et al.\ (2011), one can distinguish not very compact ($x
\lesssim 0.5$) and very compact ($x \gtrsim 0.5$) neutron stars.
Values $x \geq 0.7$ are forbidden by the causality principle (e.g.\
Haensel, Potekhin \& Yakovlev 2007).

Thus, for the neutron star in the Crab, the cooling is mainly
determined by the three parameters, $f_\ell$, $\Delta M$, and $x$.
The majority of realistic models of neutron stars have $x \lesssim
0.5$ and we restrict ourselves to these models. Their cooling weakly
depends on $x$, so that we can consider the effect of $f_\ell$ and
$\Delta M$ but neglect the effect of $x$ (although we comment on the
latter below).

\begin{figure}[tc]
\includegraphics[width=\columnwidth]{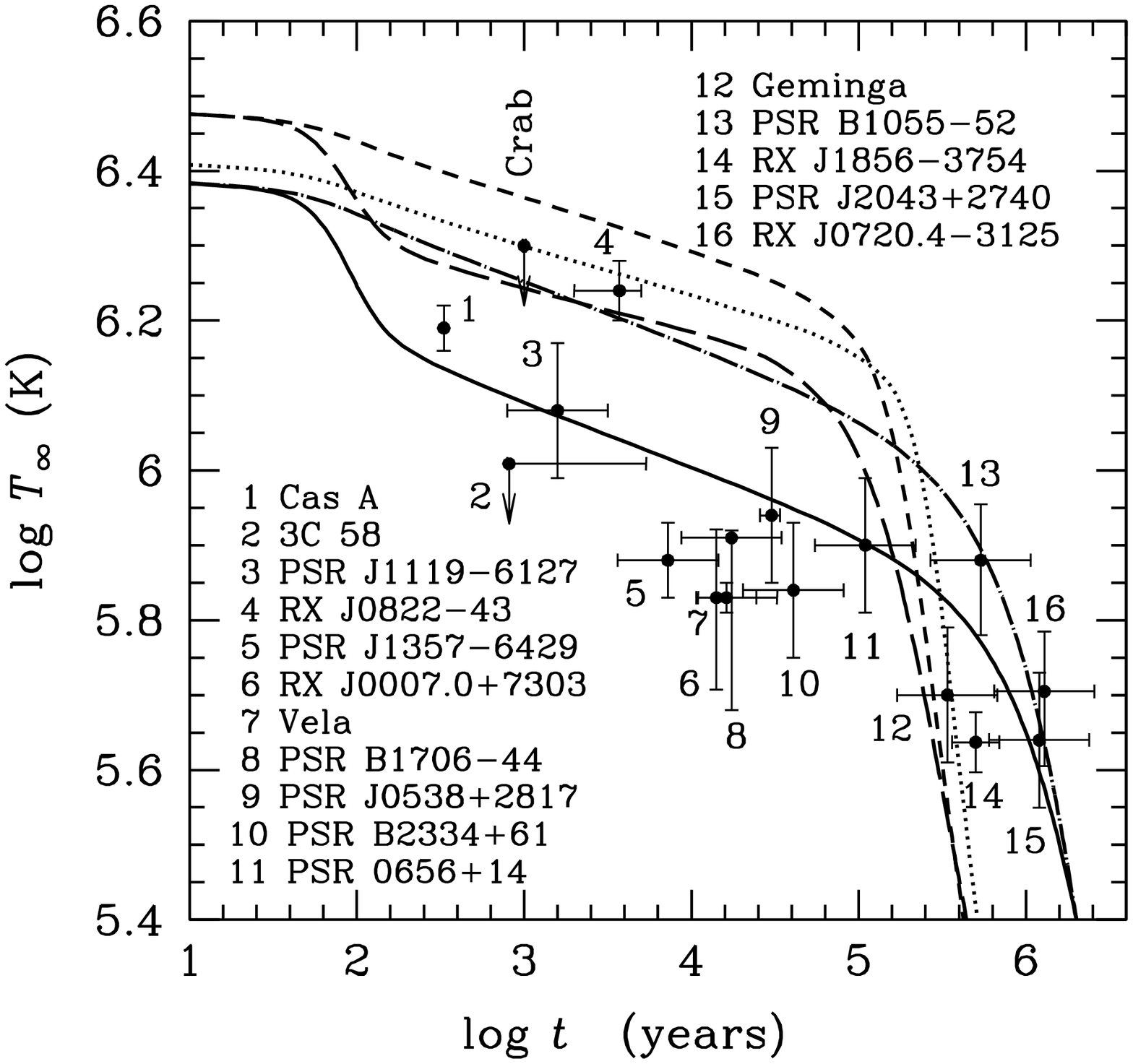}
\caption{Theoretical cooling curves for a 1.4~$M_\odot$ neutron star
with the APR EOS in the core compared with the
3$\sigma$ upper limit of $T_\infty$ for the Crab pulsar (inferred
from the blackbody fits) and with measured or constrained $T_\infty$
for some other isolated neutron stars. The solid curve is for the
standard neutrino candle with an iron heat-blanketing envelope. The
long-dashed curve is for the same star but with a fully accreted
envelope. The dot-dashed curve corresponds to a star with strong
proton superfluidity in the core and an iron heat-blanketing
envelope; the short-dashed curve is for a star with the same core but
for a fully accreted envelope; the dotted curve is for the same core
but for a partly accreted envelope with $\Delta M =
10^{-9}\,M_\odot$. (See text for details.)}
\label{fig:ts}
\end{figure}

Our new observational upper limits on $T$ are high, comparable to the highest surface temperatures which a cooling neutron star can have. Thus these $T$ limits allow the Crab pulsar to have almost any theoretically possible neutrino cooling rate (from very fast to very slow). 
It is nevertheless instructive to compare the upper limits on $T$
for the Crab pulsar with theoretical models of the warmest cooling
neutron stars. This comparison is illustrated in Fig.~\ref{fig:ts}.
The figure shows some theoretical cooling curves
($T_\infty(t)=T(t)\,\sqrt{1-x}$).
The data for stars other than the
Crab are taken from the same references as in Shternin et al.\
(2011). The cooling curves are calculated for a model of a neutron star
whose core consists of nucleons and has the EOS of Akmal,
Pandharipande \& Ravenhall (1998) (APR). Specifically, we use a
version of the APR EOS denoted as APR~I in Gusakov et al.\ (2005).
The maximum mass of a stable neutron star with this EOS is
$M_\mathrm{max}=1.929\, M_\odot$; the powerful direct Urca process of
neutrino emission is allowed in stars with $M>1.829\, M_\odot$. In
Fig.~\ref{fig:ts} we take a star with $M=1.4\,M_\odot$ ($R=12.14$
km).
The upper limit of $T_\infty$
for the Crab pulsar ($ \log T^{\rm BB}_\infty(3\sigma)
~{\rm [K]}=6.30$; $T_\infty^\mathrm{BB}\approx 2$~MK) is from the
blackbody fits at the 3$\sigma$ level for this choice of $M$ and $R$
(\S~\ref{s:bb}).

The solid line in Fig.~\ref{fig:ts} is the basic cooling curve for a
non-superfluid APR $1.4\,M_\odot$ neutron star with the iron heat
blanket ($\Delta M = 0$). This star cools via the modified Urca
process, its $f_\ell =1$, and thus it represents the standard
neutrino candle.
Its internal temperature at the Crab age would be
$\widetilde{T}_{i\rm SC}\approx 2.23 \times 10^8$~K, with the
neutrino cooling rate $\ell_\mathrm{SC} \approx 3.7 \times 10^4$
K~yr$^{-1}$.
The basic curve goes below the $T_\infty$ upper limit for the Crab
pulsar.
Therefore, the assumption that the Crab pulsar is the standard
neutrino candle is compatible with the given upper limit. This
conclusion can also be made from previous work, e.g. Kaminker et al.\
(2006).

Next we consider two cooling regulators -- the neutrino emission rate
$f_\ell$ in the stellar core (to change the internal temperature
$\widetilde{T}_i$)
and the amount $\Delta M$ of light elements in the envelope (to
change $T_\infty$ for a given
$\widetilde{T}_i$).
The long-dashed line in Fig.~\ref{fig:ts} shows cooling of the same
standard candle but with a fully accreted envelope. Light elements
increase the thermal conductivity, make the envelope more heat
transparent, and increase $T_\infty$ (for a given
$\widetilde{T}_i$).
The increase is substantial but cannot raise the cooling curve above
the $T_\infty$ upper limit for the Crab pulsar. Thus, the Crab pulsar
can well be the standard neutrino candle inside and have a
fully-accreted envelope outside.

The pulsar can also be warmed by reducing its neutrino emission below
the standard level. The dot-dashed line in Fig.~\ref{fig:ts}shows the cooling of the same star as in the previous paragraph with the iron heat-blanketing
envelope, but with strong proton superfluidity in the core. This
superfluidity greatly suppresses the modified Urca process of
neutrino emission and the bremsstrahlung emission of neutrino pairs
in proton-proton and proton-neutron collisions (and also suppresses
proton heat capacity in the core). Under these conditions, the star
cools via neutrino bremsstrahlung emission in neutron-neutron
collisions. In this scenario, the normalized neutrino cooling rate is
small, $f_\ell \approx 0.01$, and the core warmer. Exact values of
the critical temperature for onset of proton superfluidity are
unimportant here; the critical temperature in the core should be
higher than a few times of $10^9$~K to establish this very slow
cooling regime. It is thought to be one of the slowest cooling
regimes that can be realized in a cooling star (without additional
heat sources) and it produces stars with the hottest cores. The
dot-dashed line shows that this hottest star has about the same
surface temperature as the standard neutrino candle with a fully
accreted envelope; it is not forbidden by our observations.

The short-dashed curve in Fig.~\ref{fig:ts} displays cooling of the
neutron star with the same very slow cooling rate as previously but
now with a fully-accreted envelope. With respect to the basic solid
cooling curve, its surface temperature is increased by two factors:
by proton superfluidity in the core ($f_\ell \approx 0.01$) and by
accreted matter in the envelope ($\Delta M=\Delta M_\mathrm{max}$).
The large increase makes the surface exceptionally warm, with
$T_\infty$ higher than the upper limit for the Crab pulsar. Our
observational upper limit to $T_\infty$ does restrict this particular
model. Thus, if the Crab pulsar does have strong proton superfluidity
($f_\ell \approx 0.01$) it can only have a partially accreted
envelope, with $\Delta M \lesssim 10^{-9}\,M_\odot$. This is
demonstrated by the dotted cooling curve calculated for $\Delta
M=10^{-9}\,M_\odot$; it hits the observational $T_\infty$ limit.

We also considered a sequence of cooling models with progressively
weaker proton superfluidity in the Crab pulsar core. The parameter
$f_\ell$ varies then from $f_\ell\approx 0.01$ for strongly
superfluid to $f_\ell=1$ for a non-superfluid star.
At $f_\ell \gtrsim 0.15$
we find $T_\infty$ lower than the observational upper limit for any
$\Delta M$. All these models are therefore allowed by the
observations. At lower $f_\ell$ and $\Delta M \gtrsim
10^{-9}\,M_\odot$ the pulsar surface would be warmer than the
observational limit in disagreement with our observations.

Our conclusions are fairly independent of neutron star mass. Indeed
we have considered a wide range of masses from 1 to 1.8~$M_\odot$
(for stars with the APR EOS). The cooling curves stay essentially the
same, and the observational upper limits do not change much
(Figs.~\ref{f:kT_bbnorm} \& ~\ref{f:kT_nsa}). For instance, $\log
T^{\rm BB}_\infty(3\sigma)
=6.31$ from the blackbody fits for an $M=1.8 \,M_\odot$ ($R=11.38$
km) star. Moreover, the cooling curves stay nearly the same for a
large variety of EOSs of dense nucleon matter. These include the 9
original versions of the phenomenological PAL EOS (Prakash, Ainsworth
\& Lattimer 1988), as well as 3 other versions of this EOS with the
symmetry energy of nuclear matter proposed by Page \& Applegate
(1992); and the SLy EOS (Douchin \& Haensel 2001). These neutron star
models are all not very compact ($x \lesssim 0.5$) which justifies
our consideration of not very compact stars (see above). The very
weak dependence of the cooling curves on $M$, $R$ and the EOS for
standard candles and for very slowly cooling neutron stars is not new
and has been in the literature [starting from the paper by Page \&
Applegate (1992); see, e.g. Yakovlev \& Pethick (2004), for references].

Instead of the
3$\sigma$ upper limits of $T_\infty$ we could also use
less conservative 2$\sigma$ limits. For instance, we have
$\log T^{\rm BB}_\infty(2\sigma)=6.26$
($T_\infty^\mathrm{BB}(2\sigma)\approx 1.8 $~MK), for the
$1.4\,M_\odot$ neutron star with the APR EOS. In this case we obtain
$f_\ell \gtrsim 0.013$ (very slow or any faster cooling) for an iron
envelope, $f_\ell \gtrsim 1$ (standard or any faster cooling) for a
fully accreted envelope. Also, the upper limits of $T_\infty$,
inferred from neutron star atmosphere fits (\S \ref{s:nsa}), are
higher than for the black-body fits [e.g.�$\log T^{\rm NSA}_\infty(2\sigma)=6.44$ and $\log T^{\rm NSA}_\infty(3\sigma)=6.45$ for the $1.4\,M_\odot$ APR star]; they would of course be less
restrictive.

For completeness we note that massive neutron stars may become
especially compact. If they were also slow neutrino coolers, their
cooling curves would depend on the compactness parameter $x$,
Eq.~(\ref{eq:x}), and their $T_\infty$ would noticeably decrease with
increasing $x$ (at $x \gtrsim 0.5$) (Yakovlev et al.\ 2011). Such
(less realistic) models would be less restricted by the observational
$T_\infty$ limits.

Let us remark that according to the majority of current theories
massive neutron stars cool faster than the standard neutrino candles
due to the onset of fast neutrino emission in their cores. The mass
range of stars which demonstrate faster cooling is very uncertain.
Our $T_\infty$ limits do not constrain the parameters of the Crab
pulsar if it is a rapidly cooling star.

We note that our model for proton superfluidity to suppress the
neutrino emission should be considered as an example. One may use a
more general approach (Yakovlev et al.\ 2011) introducing the
normalized neutrino emission rate $f_\ell$ without specifying a
physical model of neutrino emission. The cooling equation contains
$f_\ell$, and it is $f_\ell$ that can be constrained from the
observations. Thus there could be several physical models of stellar
interior which give the same $f_\ell$, and the cooling theory itself
cannot discriminate between them.

The conclusions above follow from cooling calculations already
available in the literature (particularly, from the results of
Kaminker et al.\ 2006). We have repeated these calculations drawing
special attention to the Crab pulsar; our Fig.~\ref{fig:ts} is
similar to the right panel of Fig.\ 2 of Kaminker et al.\ (2006).
Notice, that our short-dashed curve goes higher than the analogous
curve in Kaminker et al.\ (2006). This is both because we assume a
fully-accreted heat-blanketing envelope $\Delta M \sim
10^{-7}\,M_\odot$, while Kaminker et al.\ (2006) took $\Delta M \sim
10^{-8}\,M_\odot$, and because we employ stronger proton
superfluidity in the core. This difference of cooling curves reflects
the uncertainty of the present cooling theory but does not affect our
principal conclusions. Finally, we emphasize that at the present
stage it would be better to use the model-independent formulation of
the cooling theory (Yakovlev et al.\ 2011), introducing $f_\ell$
instead of employing any specific physical cooling model, especially
when interpreting data.

\clearpage

\section{Summary} \label{s:sum}

We have obtained new \cha\ data of the Crab Nebula and its pulsar.
The new data were collected in such a manner to prevent telemetry
saturation of the LETGS and thus enable efficient collection of
high-time resolution data from the pulsar. We have analyzed these
data and re-analyzed our previous observation of the phase averaged
spectrum to update spectral parameters. The updated phase-averaged
spectral parameters no longer (Paper I) indicate that the Crab
line-of-sight is under abundant in oxygen given the
abundances and cross-sections employed in the spectral fitting. 
In all our analyses, we have accounted for the contribution of
scattering by interstellar dust to the extinction of X rays in an
aperture-limited measurement~--- a consideration (often ignored) in
spectral analysis of point sources observed with \cha 's exceptional
angular resolution. Here we have measured, for the first time, the
magnitude of that extinction in the direction of the Crab pulsar.

In addition, we have measured with a high precision the spectrum of
the Pulsar as a function of pulse phase and at {\it all} pulse
phases. We find highly significant variation of the power law
spectral index as a function of phase and have discovered an unusual
behavior of the spectral index as the pulse rises out of pulse
minimum on its approach towards the peak of the primary pulse.
Interestingly, this behavior appears to be connected to a similar
feature in the variation of the optical polarization as a function of
pulse phase as well as variations of the gamma-ray spectral index. In
both slot- and outer-gap models for phase-resolved radiation from the
Crab, the X-ray emission comes from synchrotron radiation of
secondary pairs. The variations in X-ray spectral index are thus
mapping the variations in pair spectrum with phase, although neither
of these models currently includes the physical elements that produce
the observed spectral variations. Therefore, the more accurate
measurements presented in this paper will be a challenge to future
modeling, and they have the hope of helping us understand the pair
cascade processes in pulsar magnetospheres.

We also use the spectral data to obtain new and more precise upper limits to the surface temperature of the neutron
star for two different models of the star's atmosphere.

We have commented on the differences in measured parameters subsequent to analyzing the same data with different releases of the response functions. 
Our experience emphasizes the importance
of accounting for the uncertainties in the response functions when
analyzing data. 
One might estimate the level of those uncertainties
by noting differences in spectral parameters using the old
(Paper I) and new (this paper) response functions, however, this might be too extreme as the newer response functions are a product of several proven refinements. 
Perhaps then these differences can serve as estimators of upper limits to the variations. 
We urge the various observatories to provide users with response functions with  errors and the tools use them. 

Finally, we clarify the means by which the observational data as to
the thermal emission may be connected to theories of neutron star
cooling and neutron star structure. Our principal conclusions, only
slightly dependent on the EOS of the pulsar core, pulsar mass, and
pulsar radius, are:
\begin{itemize}

\item
Our upper limits to the surface temperature $T_\infty$ of the
Crab pulsar weakly restrict the normalized neutrino emission rate $f_\ell$
(in units of standard candles) in the pulsar core and the amount of
light elements $\Delta M$ in the heat-blanketing envelope.

\item
Our observations allow the pulsar to have a neutrino emission rate
$f_\ell \gtrsim 0.15$ (1/6 of the standard neutrino cooling or higher
for the fastest cooling) for any amount of light elements in the
blanketing envelope. For lower neutrino emission rates from $\sim
0.15$ to $0.01$ (the lowest rate on physical grounds, e.g. due to
strong proton superfluidity in the core), our observations constrain
the pulsar to have only a limited mass of accreted material (with
$\Delta M \lesssim 10^{-9}\,M_\odot$ at $f_\ell \sim 0.01$).

\end{itemize}

The absence of strong restrictions on the properties of the Crab
pulsar follow from the current 3$-\sigma$ upper limit on $T_\infty$. Nevertheless,
this state of affairs has its own advantage. 
There is still a chance
that the Crab pulsar is {\it warm}, with the surface temperature
$T_\infty$ only slightly below the present upper limit. 
While our upper limit is not very restrictive, a {\it real measurement} of
the surface temperature just below the present upper limit would be
more restrictive, indicating that the Crab pulsar is one of the warmest neutron stars. For instance, if the temperature $T_\infty =
1.6 \times 10^6$~K ($\log T_\infty=6.20$) were measured, we would
have $f_\ell \approx 0.06$ (slow cooling) for an iron envelope,
$f_\ell \approx 10$ (ten times faster than the standard cooling) for
a fully accreted envelope, and generally $0.06 \lesssim f_\ell
\lesssim 10$ for a partially accreted envelope. Thus we would be able
to constrain the neutrino emission rate within two orders of
magnitude. This large uncertainty comes from the unknown mass of the
accreted matter. If $\Delta M$ were known, $f_\ell$ would be even
better constrained.

\acknowledgements{We acknowledge our tremendous debt to Leon Van
Speybroeck for his remarkable contributions to the development of the
\cha\ optics, to Harvey Tananbaum for his superb stewardship of the
\cha\ X-ray Center, and to Steve Murray and Michael Juda for their
support in successfully configuring the HRC shutter for this
measurement. DY acknowledges support by: the Russian Foundation for
Basic Research, grant 11-02-00253a; Rosnauka, grant NSh 3769.2012.2;
and Ministry of Education and Science of the Russian Federation,
contract 11.G34.31.001 with SPbSPU and Leading Scientist
G.~G.~Pavlov.}

\end{document}